\documentstyle[preprint,prb,aps,]{revtex}
\tightenlines
\begin{document}
\draft
\title{Optical investigation on the electronic structures of Y$_{2}$Ru$_{2}$O$_{7}$%
, CaRuO$_{3}$, SrRuO$_{3}$, and Bi$_{2}$Ru$_{2}$O$_{7}$}
\author{J. S. Lee, Y. S. Lee, and T. W. Noh}
\address{School of Physics and Research Center for Oxide Electronics, Seoul National\\
University, Seoul 151-747, Korea}
\author{K. Char, Jonghyurk Park, and S.-J. Oh}
\address{School of Physics and Center for Strongly Correlated Materials\\
Research, Seoul National University, Seoul 151-747, Korea}
\author{J.-H. Park}
\address{Department of Physics and electron Spin Science Center, Pohang University of%
\\
Science and Technology, Pohang 790-784, Korea}
\author{C. B. Eom}
\address{Department of Materials Science and Engineering, University of\\
Wisconsin-Madison, Madison, WI 53706, USA}
\author{T. Takeda and R. Kanno}
\address{Department of Chemistry, Kobe University, Hyogo 657, Japan}
\date{\today }
\maketitle

\begin{abstract}
We investigated the electronic structures of the bandwidth-controlled
ruthenates, Y$_{2}$Ru$_{2}$O$_{7}$, CaRuO$_{3}$, SrRuO$_{3}$, and Bi$_{2}$Ru$%
_{2}$O$_{7}$, by optical conductivity analysis in a wide energy region of 5
meV $\sim $ 12 eV. We could assign optical transitions from the systematic
changes of the spectra and by comparison with the O 1$s$ x-ray absorption
data. We estimated some physical parameters, such as the on-site Coulomb
repulsion energy and the crystal-field splitting energy. These parameters
show that the 4$d$ orbitals should be more extended than 3$d$ ones. These
results are also discussed in terms of the Mott-Hubbard model.
\end{abstract}


\newpage

\section{Introduction}

The Mott transition has been widely studied in strongly correlated electron
systems as one of the most important mechanisms for metal-insulator
transition.\cite{Mott} For 3$d$ transition metal oxides, which have rather
localized electron orbitals, the electron-electron ($el$-$el$) correlation
becomes significant and plays important roles in determining their physical
properties. Up to now, most works on the Mott transition have been focused
on the 3$d$ transition metal oxides, especially with early transition
metals, such as Ti and V.\cite{RMP 1998} On the other hand, for 4$d$ and 5$d$
transition metal oxides, it has been known that the $el$-$el$ correlation
becomes weaker due to their more extended $d$ orbitals.

Since unconventional superconductivity was observed for Sr$_{2}$RuO$_{4}$, 
\cite{Maeno} there has been large interest in ruthenates. Although
ruthenates belong to the 4$d$ transition metal oxides, some experimental
works have suggested that the correlation effects play important roles. By
investigating optical conductivity spectra $\sigma (\omega )$ of SrRuO$_{3}$%
, Kostic {\it et al.} claimed that its electrodynamics could not be
explained by the Fermi-liquid theory.\cite{Schlesinger} Lee {\it et al.}
reported observation of pseudogap formation in BaRuO$_{3}$, which has some
similarities with the pseudogaps of high T$_{c}$ superconductors.\cite{YSLEE}
From photoelectron spectroscopy (PES) and x-ray absorption (XAS)
measurements of SrRuO$_{3}$, Fujioka {\it et al.} reported that its electron
correlation effects should be considered to be important.\cite{Theory U}

Electrical properties of the ternary ruthenates show systematic changes from
a Mott insulator to a band metal. Y$_{2}$Ru$_{2}$O$_{7}$ has the pyrochlore
structure and shows an insulating behavior. Both CaRuO$_{3}$ and SrRuO$_{3}$
have the perovskite structures and show metallic responses. Note that SrRuO$%
_{3}$ is more metallic than CaRuO$_{3}$, and that SrRuO$_{3}$ becomes
ferromagnetic below 160 K. Bi$_{2}$Ru$_{2}$O$_{7}$ has the pyrochlore
structure and behaves as a Pauli paramagnetic metal. All of the above
pyrochlore and perovskite ruthenates have four 4$d$ electrons in the Ru 4+
ions with the low spin configuration. Since all of the ruthenates have
three-dimensional networks of corner-shared RuO$_{6}$ octahedra, they are
expected to have similar electronic structures. Cox {\it et al.}
investigated systematically the electronic structures of the ruthenates
using PES and electron energy loss spectroscopy.\cite{Cox} Their results
were consistent with a progressive decrease in the transfer-energy integral
that determines the Ru 4$d$ bandwidth in the series Y$_{2}$Ru$_{2}$O$_{7}$ 
%TCIMACRO{\TEXTsymbol{<} }%
%BeginExpansion
\mbox{$<$}%
%EndExpansion
CaRuO$_{3}$ 
%TCIMACRO{\TEXTsymbol{<} }%
%BeginExpansion
\mbox{$<$}%
%EndExpansion
SrRuO$_{3}$ 
%TCIMACRO{\TEXTsymbol{<} }%
%BeginExpansion
\mbox{$<$}%
%EndExpansion
Bi$_{2}$Ru$_{2}$O$_{7}$, and the insulating behavior of Y$_{2}$Ru$_{2}$O$%
_{7} $ could be understood in terms of the correlation-induced electron
localization, i.e. the Mott transition.

An optical spectroscopy measurement in a wide spectral region is also known
to be a powerful tool to analyze the electronic structures of some highly
correlated electron systems.\cite{Manganites U} In this paper, we
investigated the electronic structures of the ternary ruthenates, Y$_{2}$Ru$%
_{2}$O$_{7}$, CaRuO$_{3}$, SrRuO$_{3}$, and Bi$_{2}$Ru$_{2}$O$_{7}$ by
measuring reflectivity spectra from 5 meV to 30 eV. Their $\sigma (\omega )$
showed systematic changes as the Ru 4$d$ bandwidth increased. Combined with
the O 1$s$ XAS data, we could determine optical transitions of ruthenates.
We found that the old assignments of the transitions by one of us (T.W.N.), 
\cite{J. S. Ahn} related to the Hubbard bands and the quasi-particle (QP)
band of the perovskite ruthenates, were not correct. Such a mistake occurred
because the assignments were made solely on the basis of $\sigma (\omega )$
below 5 eV. However, in this study, the transition assignments were made
based on $\sigma (\omega )$ of four different compounds in a much wider
spectral region. And, the assignments were also consistent with the O 1$s$
XAS data. From the fitting on $\sigma (\omega )$ of CaRuO$_{3}$, we could
estimate some physical quantities, such as the on-site Coulomb repulsion
energy $U$ and the crystal field splitting 10$Dq$.

\section{Experimental Techniques}

Optical spectra of the perovskite ruthenates were measured using epitaxial
films. The SrRuO$_{3}$ and the CaRuO$_{3}$ films were epitaxially grown on
the SrTiO$_{3}$ substrates using the pulsed laser deposition and sputtering
techniques, respectively.\cite{SrRuO3,Eom} The film thicknesses were larger
than 4000 \r{A}, which are much thicker than the penetration depth of the
light, so the reflected light from the SrTiO$_{3}$ substrates became
negligible. For the CaRuO$_{3}$ films, it is well known that their transport
properties can be changed significantly, namely from a metal to an
insulator, by strain effects.\cite{Eom} To obtain films whose properties
were similar to those of a bulk sample, we used a substrate with a large
miscut angle of 5 $\sim $ 7 $^{\circ }$.

Optical spectra of the pyrochlore ruthenates were measured using high
density polycrystalline samples. The Bi$_{2}$Ru$_{2}$O$_{7}$ and Y$_{2}$Ru$%
_{2}$O$_{7}$ polycrystalline samples were synthesized by sintering under a
high pressure of around 3 GPa.\cite{Takeda2} Such a high pressure was
applied to obtain high density samples, since it was difficult to obtain
reliable optical data using samples with high porosity. Since the pyrochlore
phase is a cubic phase, the optical constants of the pyrochlore ruthenates
should be isotropic. So, we could determine their optical constants even
from polycrystalline samples. Before optical measurements, the surfaces of
the polycrystalline samples were polished up to 0.3 $\mu $m. After optical
measurements, thin gold films were evaporated on the samples and used to
make corrections due to scattering from rough sample surfaces.

We measured reflectivity spectra from 5 meV to 30 eV using numerous
spectrophotometers at room temperature. In the energy region between 5 meV
and 0.6 eV, we used a conventional Fourier transform spectrophotometer.
Between 0.5 eV and 6.0 eV, we measured the spectra using a grating
spectrophotometer. In the deep ultraviolet region above 6.0 eV, we used the
synchrotron radiation from the normal incidence monochromator beam line at
Pohang Accelerator Laboratory. From the reflectivity spectra in a wide
region, we used the Kramers-Kronig (K-K) transform to obtain $\sigma (\omega
)$. For this analysis, the reflectivity below 5 meV was extrapolated with
the Hagen-Rubens relation for the metallic ruthenate samples and with a
constant value for the insulating Y$_{2}$Ru$_{2}$O$_{7}$ sample. For a high
frequency region, the reflectivity value at 30 eV was used for
reflectivities up to 40 eV, above which $\omega ^{-4}$ dependence was
assumed.\ We found that the results presented in this paper were nearly
independent of the details of the extrapolation. To check the validity of
our K-K analysis, we also independently determined $\sigma (\omega )$ using
spectroscopic ellipsometry techniques in the photon energy range of 1.5 $%
\sim $ 5.5 eV. $\sigma (\omega )$ data from the spectroscopic ellipsometry
agreed quite well with the results from the K-K analysis.\cite{KK}

\section{Results and analysis}

Figure 1 shows room temperature $\sigma (\omega )$ of Y$_{2}$Ru$_{2}$O$_{7}$%
, CaRuO$_{3}$, SrRuO$_{3}$, and Bi$_{2}$Ru$_{2}$O$_{7}$. As displayed in
Figs. 1(a) and (b), the spectra of Y$_{2}$Ru$_{2}$O$_{7}$, CaRuO$_{3}$, and
SrRuO$_{3}$ show similar interband transitions in the energy region between
3 and 12 eV. Three strong peaks around 3, 6, and 10 eV are commonly
observed. These facts indicate that the electronic structures of these
compounds should be quite similar. However, $\sigma (\omega )$ of Bi$_{2}$Ru$%
_{2}$O$_{7}$ looks different from those of other ruthenates. While the peak
around 10 eV disappears, the peak around 6 eV is much enhanced.
[Nevertheless, as we will show later, these features also can be understood
in terms of an electronic structure which is basically similar to those of
other ruthenates, and the differences in the spectral distribution come from
the differences in the allowed transitions.]

It should be noted that the spectra in the energy region below 2.5 eV vary
from a sample to a sample. As shown in Fig. 1(a), $\sigma (\omega )$ of Y$%
_{2}$Ru$_{2}$O$_{7}$ does not have any Drude-like peak, in agreement with
the fact that it is in an insulating state. Also, a rather weak peak appears
around 1.7 eV. As shown in Fig. 1(b), $\sigma (\omega )$ of CaRuO$_{3}$ has
a coherent Drude-like peak and an incoherent weak peak around 1.7 eV, which
is quite similar to that observed in Y$_{2}$Ru$_{2}$O$_{7}$. On the other
hand, $\sigma (\omega )$ of SrRuO$_{3}$ and Bi$_{2}$Ru$_{2}$O$_{7}$ have the
Drude-like peaks, but do not show any peak around 1.7 eV. Since the
ruthenates become more metallic as it goes from Y$_{2}$Ru$_{2}$O$_{7}$ to Bi$%
_{2}$Ru$_{2}$O$_{7}$, the systematic changes of the coherent Drude-like peak
and the incoherent peak around 1.7 eV could be closely related to electrical
properties of the systems.

Figure 2 is a schematic diagram for the electronic structure of CaRuO$_{3}$.
The diagram is mainly composed of three parts: (1) the Ru 4$d$ states near $%
E_{F}$, (2) the O 2$p$ states around 5 eV below $E_{F}$, and (3) the
unoccupied Ca 3$d$ states around 6 eV above $E_{F}$. CaRuO$_{3}$ is a
metallic compound very close to the Mott transition.\cite{Cox} According to
the dynamic mean field theory,\cite{RMP 1996} for a metallic sample near the
Mott transition, its one particle spectral function will be split into two
Hubbard bands in addition to the QP band located at $E_{F}$. Therefore, the
Ru 4$d$ states might be split into upper and lower Hubbard bands by the
on-site Coulomb repulsion energy $U$ with the QP band at $E_{F}$. And, the
upper Hubbard band might be further splits into states of the $t_{2g}$ and
the $e_{g}$ symmetries by 10$Dq$.

To explain the electronic structures of other compounds, we should make two
minor modifications in Fig. 2. First, the unoccupied Ca 3$d$ states should
be replaced by the unoccupied Y 4$d$, Sr 4$d$, or Bi 6$p$ states. For the Y 4%
$d$ and Bi 6$p$ states, they might be located at energies lower than those
of the Ca 3$d$ and Sr 4$d$ states and could overlap with the $e_{g}$ states. 
\cite{Okamoto,J. H. Park} Second, the details of the Ru 4$d$ states near $%
E_{F}$ should vary according to the $el$-$el$ correlation effect. For
insulating Y$_{2}$Ru$_{2}$O$_{7}$, the Hubbard bands are completely split
due to the large $el$-$el$ correlation, and the QP peak disappears.\cite{Cox}
On the other hand, for a band metal, such as Bi$_{2}$Ru$_{2}$O$_{7}$, the $%
el $-$el$ correlation becomes quite weak, and the Hubbard bands merge to
form a single partially filled band at $E_{F}$. For SrRuO$_{3}$, the
structures of the Ru 4$d$ states might be close to those for Bi$_{2}$Ru$_{2}$%
O$_{7}$, because SrRuO$_{3}$ is known to be less correlated than CaRuO$_{3}$%
. \cite{Cox} [Note that, with such changes, the diagram in Fig. 2 becomes
consistent with the results of the PES and the XAS experiments on SrRuO$_{3}$%
\cite{Okamoto} and (Bi,Y)$_{2}$Ru$_{2}$O$_{7}$. \cite{J. H. Park}]

To understand $\sigma (\omega )$ of CaRuO$_{3}$, we considered all of the
possible optical transitions, which were marked as solid lines at the top of
the electronic structure diagram in Fig. 2. According to the Fermi's golden
rule,\cite{Fermi-golden} the optical transition rate from an initial state $%
|i\rangle $ to a final state $|f\rangle $ can be written as

\begin{equation}
I_{i\rightarrow f}\sim |\langle f|\vec{r}_{o}\cdot \vec{p}|i\rangle
|^{2}\rho _{i}\rho _{f}\text{,}
\end{equation}
where $\rho _{i}$ and $\rho _{f}$ are densities of states for $|i\rangle $
and $|f\rangle $, respectively. Inside the transition matrix element, $\vec{r%
}_{o}$ is the space unit operator, and $\vec{p}$ is the momentum operator.
For an isolated atom, the transition matrix element allows electric dipole
transitions with the selection rule of $\Delta L=\pm 1$, where $\Delta L$
means the difference between the angular momentum quantum numbers for the
state $|i\rangle $ and $|f\rangle $. For solids, the matrix element also
depends on the overlap integral between $|i\rangle $ and $|f\rangle $
states. Since the O 2$p$ bands are hybridized with the Ru 4$d$ and the Ca 3$d
$ bands, the transitions from the O 2$p$ to the Ru $t_{2g}$ (Transition A),
the Ru $e_{g}$ (Transition B), and the Ca 3$d$ states (Transition C) will be
allowed. The transitions between the Ru 4$d$ states should be forbidden in
an isolated atom. However, in real oxides, some interatomic transitions
could be allowed, and some intraatomic transition might be possible due to
the hybridization with the O 2$p$ bands and/or local distortions.\cite
{Manganites U} [Note that the transitions between the Ru 4$d$ states should
be much weaker than the charge transfer transitions from the O 2$p$ to the
Ru 4$d$ states.] For CaRuO$_{3}$, three transitions from the occupied Ru $%
t_{2g}$ states to the unoccupied Ru $t_{2g}$ (Transition $\alpha $), the Ru $%
e_{g}$ (Transition $\beta $) and the Ca 3$d$ states (Transition $\gamma $)
were considered. [Other transitions involving the QP states were possible,
but found to be quite small except for the Drude peaks for metallic samples.]

Interestingly, most of the expected interband transitions are observed in $%
\sigma (\omega )$ of CaRuO$_{3}$. As shown in Fig. 3, $\sigma (\omega )$ of
CaRuO$_{3}$ is composed of three strong peaks around 3, 6, and 10 eV, and
two weak peaks around 1.7 and 4.3 eV. The large differences in peak
strengths might be originated from the different characters of each
excitation. We fitted $\sigma (\omega )$ of CaRuO$_{3}$ below 12 eV with a
Drude model and five Lorentz oscillators, whose contributions are
represented by the dotted lines.\cite{Drude discussion} The functional forms
of the Lorentz oscillators are

\begin{equation}
\sigma (\omega )=\sum_{i}\frac{S_{i}\Gamma _{i}\omega _{i}^{2}}{(\omega
^{2}-\omega _{i}^{2})^{2}+\Gamma _{i}^{2}\omega _{i}^{2}}\text{,}
\end{equation}
where $S_{i}$, $\Gamma _{i}$, and $\omega _{i}$ represent the strength, the
damping constant, and the frequency of the $i$th Lorentz oscillator,
respectively. The values of the fitting parameters are listed in TABLE I.
Note that the oscillator strengths of $i$=2, 4, and 5 are stronger by about
a factor of 5 than those of $i$=1 and 3.

To assign the interband transitions of CaRuO$_{3}$ properly, we found that
it was very helpful to check $\sigma (\omega )$ of Y$_{2}$Ru$_{2}$O$_{7}$.
As shown in Fig. 1(a), the first and the second peaks appear at 1.7 eV and
3.2 eV, respectively, which are similar to the low lying peak positions of
CaRuO$_{3}$. Considering that Y$_{2}$Ru$_{2}$O$_{7}$ is a Mott insulator, we
assigned the lowest peak as the transition between the Hubbard bands.
Actually, similar features were also observed in other insulating
ruthenates, such as Ca$_{2}$RuO$_{4}$ and $c$-axis Sr$_{2}$RuO$_{4}$.\cite
{Puchukov,Tokura Ru214} Thus, it is quite reasonable that the oscillators of 
$i$=1 and 2 for CaRuO$_{3}$ should be assigned as Transitions $\alpha $ and
A, respectively. Considering the energy differences between each excitation
and their relative strengths, we assigned the oscillators of $i$=4 and 5 as
Transitions B and C, respectively. Our fitting also shows that there is
another weak oscillator around 4.36 eV, which was assigned as Transition $%
\beta $. Note that this transition should have much lower strength than any
other transitions due to the selection rule, and that this excitation could
not be seen in $\sigma (\omega )$ of other ruthenates.

Our peak assignments of CaRuO$_{3}$ are also consistent with the results of
the O 1$s$ XAS experiments, which probe the unoccupied states of the Ru 4$d$
and the Ca 3$d$ orbitals which are strongly hybridized with the O 2$p$
orbitals. As shown in the inset of Fig. 3, three peaks were observed around
530, 533, and 537\ eV. The near edge intensities around 530 and 533 eV are
attributed to the states with the mixed O 2$p$-Ru 4$d$ characters of the $%
t_{2g}$ and the $e_{g}$ states, respectively. And, the peak around 537 eV
comes from the Ca 3$d$ states. The peak positions of the Ru $e_{g}$ and the
Ca 3$d$ states relative to the unoccupied Ru $t_{2g}$ states are about 3 and
7 eV, respectively. These energy differences are consistent with those
between the optical $p$-$d$ transitions A and B (3.14 eV), and A and C (6.47
eV), respectively. This indicates that our peak assignments are quite
reasonable.

Based on the assignments for $\sigma (\omega )$ of CaRuO$_{3}$, $\sigma
(\omega )$ of the other ruthenates can be assigned in similar ways. For Y$%
_{2}$Ru$_{2}$O$_{7}$, the distinct peaks near 3 and 6 eV can be assigned as
Transitions A and B, respectively. And, the peak around 9 eV can be assigned
as Transition C from the O 2$p$ states to the Y 4$d$ states. For SrRuO$_{3}$%
, Transitions A and B can be assigned in a similar way, and Transition C
from the O 2$p$ states to the Sr 4$d$ states appears around 10 eV. On the
other hand, these peak values of the $p$-$d$ transitions are smaller by
about 3.0 eV than the energy difference between the corresponding states
whose positions were measured by PES and XAS.\cite{Okamoto} The difference
might be due to the formation of an exciton in the optical process, whose
binding energy will lower the value of transition peak. Also, the spectral
weights lying closer to $E_{F}$ would contribute to $\sigma (\omega )$ more
significantly than those further away from $E_{F}$, so the measured value of
transition energy in $\sigma (\omega )$ could be smaller than the actual
energy separation of the initial and final states.

As shown in Fig. 1, $\sigma (\omega )$ of Bi$_{2}$Ru$_{2}$O$_{7}$ is quite
different from the other spectra. However, we can still explain its optical
spectra using an electronic structure similar to those of other ruthenates.
For Bi$_{2}$Ru$_{2}$O$_{7}$,\ the unoccupied Bi 6$p$ states, instead of the $%
d$ states for other ruthenates, are located very close to the $e_{g}$ states.
\cite{J. H. Park} While Transition B is expected near 6 eV, Transition C
between the O 2$p$ and the Bi 6$p$ states should be quite weak due to the
selection rule. On the other hand, Transition $\gamma $ from the Ru $t_{2g}$
to the Bi 6$p$ states becomes possible, and should be located at lower
energy than Transition C. Thus, the strong 6 eV structure for Bi$_{2}$Ru$_{2}
$O$_{7}$ can be attributed to Transitions B and $\gamma $, which are located
at similar energies.\cite{enhancement}

Up to this point, for some Mott-Hubbard systems, there have been
controversies about assignments of the interband transitions related to
Hubbard bands. For example, in (Sr,Ca)VO$_{3}$, the assignment of the $U$
peak, i.e. transition between the Hubbard bands, was the subject of
controversy, but was settled recently.\cite{assign CSVO1,Vanadate U1} In
ruthenates also, assignments for 1.7 and 3.0 eV transitions have been
controversial.\cite{J. S. Ahn,pd transition} Ahn {\it et al.} assigned the
3.0 eV peak as Transition $\alpha $,\cite{J. S. Ahn} but Dodge {\it et al.}
attributed the peak as a charge transfer transition from the O 2$p$ to the
Ru $t_{2g}$ states.\cite{pd transition} Based on $\sigma (\omega )$ studies
in a much wider photon energy region and the XAS studies, we finally
concluded that peaks around 1.7 and 3.0 eV should be assigned as Transitions 
$\alpha $ and A, respectively.

\section{Discussion}

From our optical analysis, we can obtain some important physical parameters
such as $U$ and 10$Dq$. The photon energy of the $d$-$d$ transition, i.e.
Transition $\alpha $, corresponds to the value of $U$. [Note that the $U$
obtained from $\sigma (\omega )$ will be smaller than the energy difference
between the upper and the lower Hubbard bands whose positions are determined
by the\ PES and XAS due to the exciton and the screening effects.] Our study
shows that the $U$ values of Y$_{2}$Ru$_{2}$O$_{7}$ and CaRuO$_{3}$ are
quite similar, i.e. about 1.7 eV. And, the energy difference between
Transitions A and B corresponds to 10$Dq$. For all of the ruthenates we
studied, the 10$Dq$ values are around 3.0 eV, which is consistent with the
theoretical results.\cite{Theory 10Dq,exchange}

It would be quite useful to compare the $U$ and 10$Dq$ values of the
ruthenates with those of other 3$d$ transition metal oxides, which were
obtained from optical measurements. The values of $U$, i.e. about 1.7 eV, in
ruthenates are much smaller than those in the 3$d$ transition metal oxides
which have similar numbers of $d$ electrons. For example, in case of
manganites with three or four electrons per Mn ion, the value of $U$ was
estimated to be larger than 3 eV.\cite{Manganites U} Instead, the value of $U
$ for CaRuO$_{3}$ is very similar to that of vanadates which have two
electrons per V ion.\cite{Vanadate U1,Vanadate U2} The small value of $U$ in
ruthenates might be due to the fact that 4$d$ electrons have more extended
wavefunctions than 3$d$ electrons. And, the value of 10$Dq$ around 3 eV in
ruthenates is larger than that of the 3$d$ transition metal oxides, such as
2.3 eV for Ni-oxides,\cite{10Dq Ni-oxides} 1.1 $\sim $ 1.8 eV for Mn-oxides. 
\cite{Mn 10Dq 1,Mn 10Dq 2} The anisotropic Coulomb potential due to the
surrounding oxygen ions could influence the 4$d$ orbitals more strongly due
to its larger size. Both the small value of $U$ and the large value of 10$Dq$
in ruthenates indicate that the 4$d$ orbitals should be more extended than 3$%
d$ ones.

It was found that the general behaviors of our spectra could be well
explained in the regime of the Mott-Hubbard model with the bandwidth
control. As the bandwidth increases, the system becomes more metallic, and
the spectral weight of the Hubbard bands transfers to the QP peak at $E_{F}$%
. Accordingly, the $d$-$d$ transitions related to the Hubbard bands, such as
Transitions $\alpha $ and $\beta $, are expected to become weaker. As it
goes from Y$_{2}$Ru$_{2}$O$_{7}$ to Bi$_{2}$Ru$_{2}$O$_{7}$, the distinct
Transition $\alpha $ in Y$_{2}$Ru$_{2}$O$_{7}$ becomes weaker and finally
disappears in SrRuO$_{3}$ and Bi$_{2}$Ru$_{2}$O$_{7}$. Especially for CaRuO$%
_{3}$, its $\sigma (\omega )$ shows the coherent peak as well as Transition $%
\alpha $, which could be a typical optical feature of a correlated metal
near the Mott transition.\cite{Vanadate U2} Simultaneously, Transition A
shifts to a lower energy by about 0.6 eV, which might be related to the
merging of the $t_{2g}$ bands to $E_{F}$ which were originally split due to
the correlation effects. These results are quite consistent with Cox {\it et
al}.'s photoemission work, which suggested that ruthenates should belong to
the class of bandwidth-controlled Mott-Hubbard systems.\cite{Cox}

\section{Summary}

We investigated optical properties of Y$_{2}$Ru$_{2}$O$_{7}$, CaRuO$_{3}$,
SrRuO$_{3}$, and Bi$_{2}$Ru$_{2}$O$_{7}$ in a wide energy region from 5 meV
to 12 eV. Their optical conductivity spectra obtained through the
Kramers-Kronig transformation show the systematic variation from a Mott
insulator of Y$_{2}$Ru$_{2}$O$_{7}$ to a band metal of Bi$_{2}$Ru$_{2}$O$_{7}
$. The general behaviors of the optical spectra could be well explained by
the Mott-Hubbard model. Combined with O 1$s$ x-ray absorption measurements,
the optical excitations were properly assigned based on the electronic
structures. From the peak assignments, we could estimate some physical
parameters, such as the on-site Coulomb repulsion energy and the crystal
field splitting energy. Compared to such parameters of the 3{\it d}
transition metal oxides, the results could describe the more extended
character of the 4$d$ orbitals in ruthenates.

\acknowledgments

We would like to thank Jaejun Yu, Yunkyu Bang, and J. Okamoto for useful
discussions. This work was supported by the Ministry of Science and
Technology through the Creative Research Initiative program, and by KOSEF
through the Center for Strongly Correlated Materials Research. The work at
POSTECH was supported by POSTECH research funds. The experiments at PLS were
supported by MOST and POSCO.

\begin{figure}[tbp]
\caption{A schematic diagram of the electronic structures of CaRuO$_{3}$.
Near $E_{F}$, the occupied and unoccupied Ru $t_{2g}$ states are located
with the QP states at $E_{F}$. There are the $e_{g}$ states above the
unoccupied $t_{2g}$ states. Outside these Ru $d$ states, there are the
occupied O 2$p$ states and the unoccupied Ca 3$d$ states. Other ruthenates,
namely, Y$_{2}$Ru$_{2}$O$_{7}$, SrRuO$_{3}$, and Bi$_{2}$Ru$_{2}$O$_{7}$,
should have similar electronic structures with a couple of differences,
which are described in the text. }
\label{Fig:2}
\end{figure}

\begin{figure}[tbp]
\caption{Detailed analysis on $\protect\sigma (\protect\omega )$ of CaRuO$%
_{3}$. The fitted results (a solid line) are well consistent with the
experimental data of $\protect\sigma (\protect\omega )$ (solid circles). The
contribution of the Drude term and five Lorentz oscillators are represented
by the dotted lines. The Lorentz oscillator at 4.36 eV is shown as a thick
dotted line for the eye. The dashed line indicates the background
contribution from higher energy peaks above 12 eV which are not shown in
this figure. The inset shows the experimental result of the O 1$s$ x-ray
absorption spectroscopy on CaRuO$_{3}$.}
\label{Fig:3}
\end{figure}

\begin{table}[bp]
\caption{Values of Lorentz oscillator fitting parameters for CaRuO$_{3}$. }
\begin{tabular}{ccccc}
& $\omega _{i}$ (eV) & $\Gamma _{i}$(eV) & $S_{i}$ ($\Omega ^{-1}$cm$%
^{-1}\cdot $ eV) & Assignments \\ 
\tableline$i$=1 (Transition $\alpha $) & 1.71 & 1.30 & 658 & $t_{2g}$ $%
\rightarrow $ $t_{2g}$ \\ 
\tableline$i$=2 (Transition A) & 3.08 & 1.82 & 2500 & O 2$p$ $\rightarrow $ $%
t_{2g}$ \\ 
\tableline$i$=3 (Transition $\beta $) & 4.36 & 1.57 & 661 & $t_{2g}$ $%
\rightarrow $ $e_{g}$ \\ 
\tableline$i$=4 (Transition B) & 6.22 & 3.22 & 3950 & O 2$p$ $\rightarrow $ $%
e_{g}$ \\ 
\tableline$i$=5 (Transition C) & 10.55 & 2.20 & 2114 & O 2$p$ $\rightarrow $
Ca 3$d$%
\end{tabular}
\end{table}


\begin{references}
\bibitem{Drude discussion}  Actually, the coherent part in CaRuO$_{3}$
cannot be fitted with the simple Drude model. This feature has many
similarities with the case of SrRuO$_{3}$. [See Ref. 4.] Since the coherent
contribution and Transition $\alpha $\ are well separated in $\sigma (\omega
)$ of CaRuO$_{3}$, the fitting for the coherent part with simple Drude model
does not influence significantly the determination of Transition $\alpha $.
\bibitem{enhancement}  XAS data of Bi$_{2}$Ru$_{2}$O$_{7}$, in Ref. 16, show
that the density of the $e_{g}$ states becomes large, which could also make
the oscillator strength of Transition B strong.
\bibitem{exchange}  The experimental and theoretical studies on SrRuO$_{3}$
reported that the exchange energy is around 0.5 and 0.65 eV, respectively.
[Refer to Ref. 8 and 25.] The comparably large 10$Dq$ and the small exchange
energy guarantee that the electrons in the Ru ion show the low spin
configuration.
\end{references}
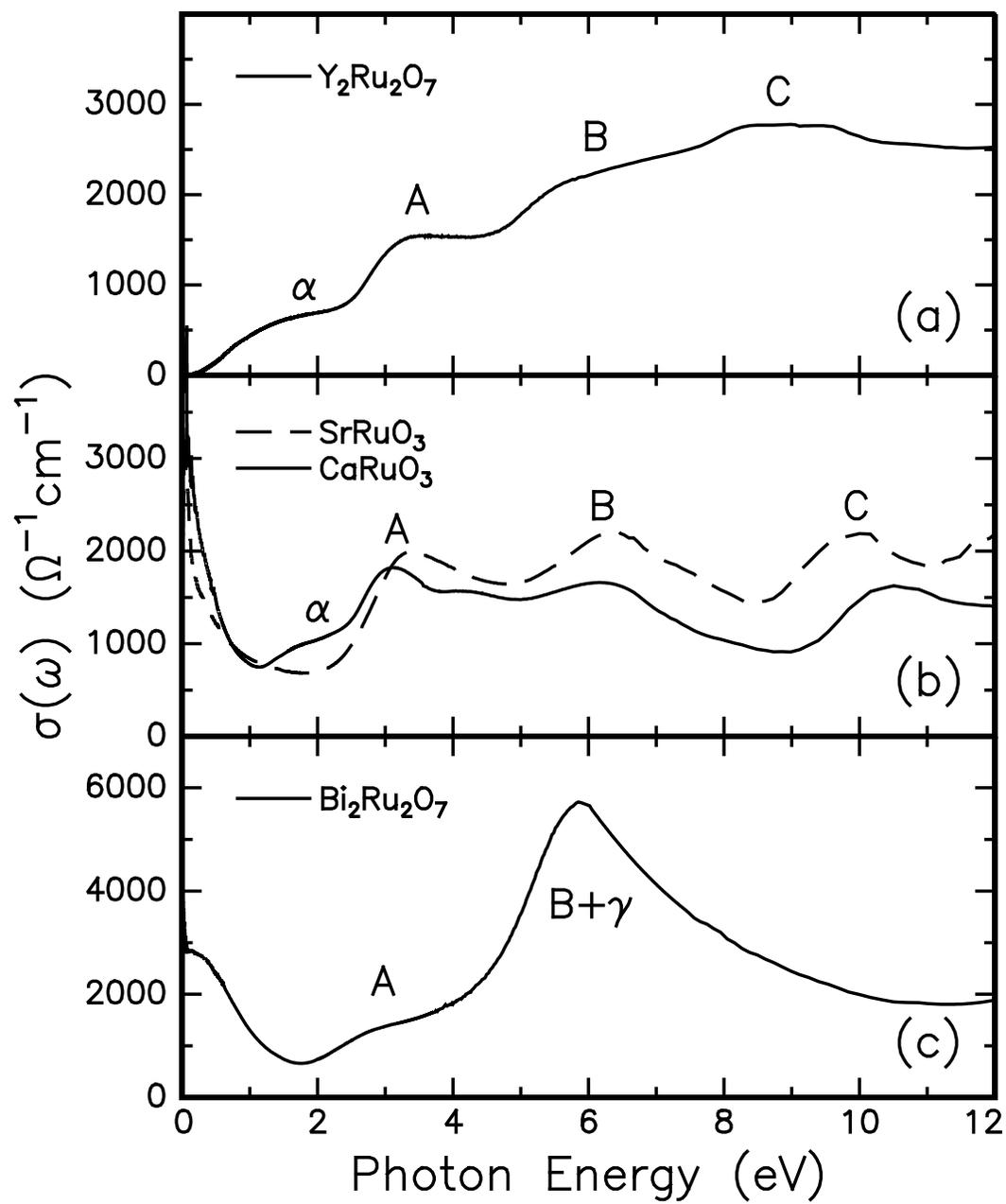
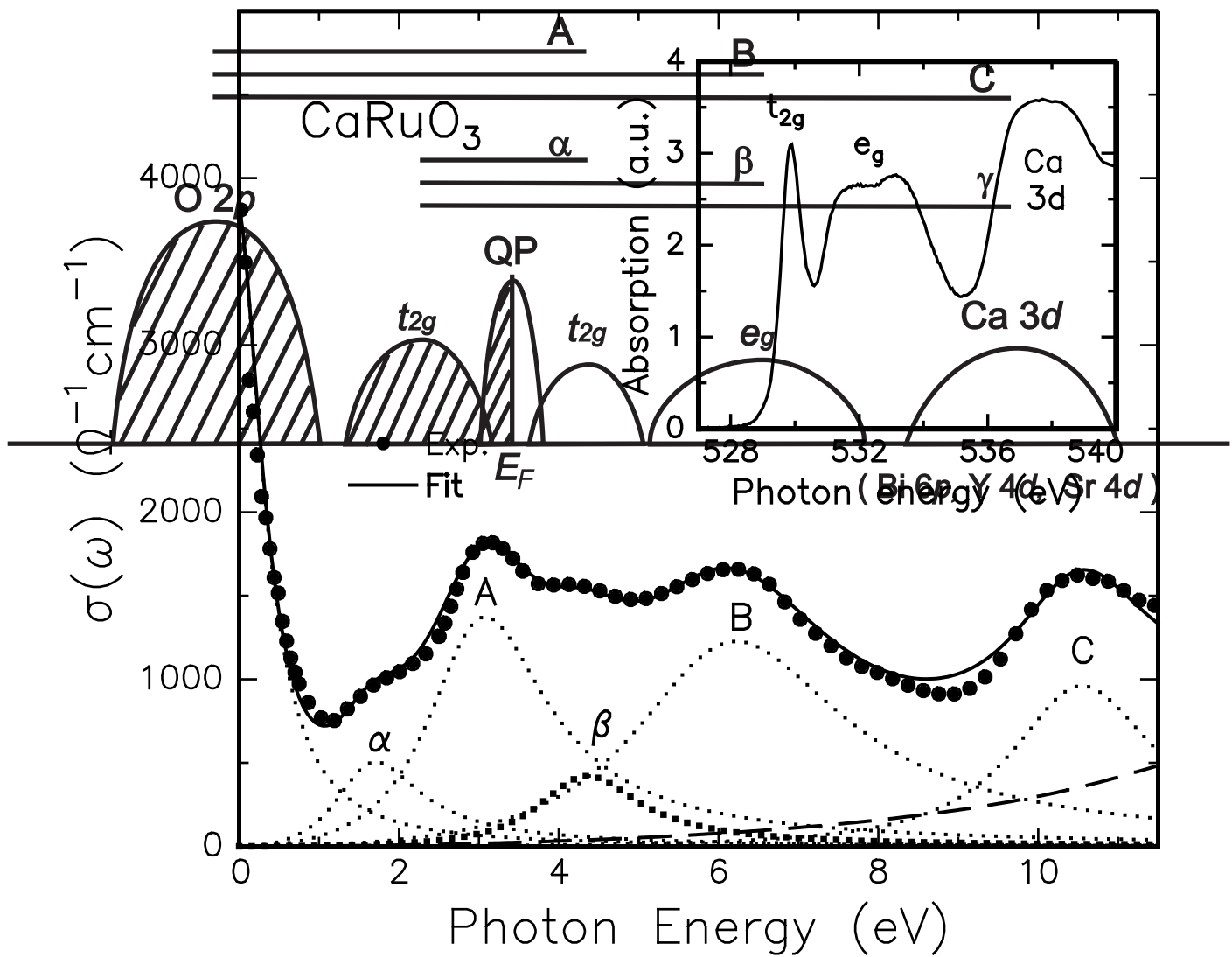
\end{document}